\documentclass{article}
\usepackage{amsmath}

\input{tcilatex}

\begin{document}

\title{D2 brane as the wormhole and the number of the universes}
\author{Pawe\l\ Gusin \\
Wroc\l aw University of Technology, Poland,\\
Faculty of Technology and Computer Science, \\
}
\date{}
\maketitle

\begin{abstract}
We construct wormhole-like solutions in the type IIA string theory. These
solutions represent wormholes in four-dimensions and are given by the D2
branes within appropraited backgrounds fields. We present the conditions on
these fields which lead to the four-dimensional wormholes. In the special
case we show how the particular solution in the type IIA theory leads to the
dynamic wormhole. We also speculate about the number of universes and the
cosmological constant.

\begin{description}
\item[PACS] : 04.50.Gh; 04.20.Gz

\item[Keywords] : wormholes, D-branes
\end{description}
\end{abstract}

\section{Introduction}

Generally, a wormhole is constructed by imposing the geometrical requirement
on spacetime that there exists a throat but no horizon. Thus the wormholes
can be interpreted as connections between different universes or topological
handles between distant parts of the same universe. Hence the spacetime with
the wormholes has the nontrivial topology. The conception of the wormhole
was first introduced by Einstein and Rosen in 1935 [1]. Moreover there are
proposals how to distinguish ( astrophysical) black holes from wormholes [2,
3]. It is known that wormholes usually cannot occur as classical solutions
of gravity due to violations of the energy conditions [4-8] thus the
non-exotic matter produces only spacetime with trivial topology.

It is known that the source of backgrounds which have geometries with
nontrivial topology are space-time extended objects called branes. Hence one
can try to obtain the wormhole solution in such backgrounds. Among the
branes are D-branes which are the degrees of freedom in string theory. On
the classical level a D-brane is a submanifold $\mathcal{M}$ on which the
ends of the string terminated. Hence the D-brane corresponds to the Dirchlet
boundary conditions that have been put on the open string. The dimensions of
the D-branes depend on the string theory considered.\ From the other side
the branes are solutions of the low energy approximation of the string
theory by\ a field theory (supergravity). In this approximation the
equations of motion involve fields from Neveu-Schwarz (NSNS) and
Ramond-Ramond (RR) sectors: the graviton, the dilaton and the other various
antisymmetric tensor fields. The solutions of these equations form
backgrounds in which the string and D-branes are propagating. In the
supergravity approximation of the string theory the evolution of a D-brane
is described by the non-linear Dirac-Born-Infeld (DBI) action which consists
of the background fields which are pulled-back on the world-volume of the
D-brane. Hence the evolution of such D-branes depend on these fields and the
internal geometry of the D-brane. Thus the D-brane is propagating in the
background of branes and fields which are field theoretic solutions. We ask:
can the equations governing the D-brane evolution be interpreted as the
equations for the wormhole in four dimensions? We will show that such an
interpretation is possible for some backgrounds. As a result we will obtain
the Lorentzian wormhole. In string theory the Euclidean wormholes were
considered in [9-12].

The aim of this paper is to the relate four-dimensional wormholes to the
solutions of string theory. We present backgrounds with branes which lead to
the wormholes in four dimensional spacetime.

In section 2 we recall the construction of the wormholes and equations of
motion. In section 3 we reproduce these equations from DBI action and
consider a type IIA string background with a D2 brane. This D2 brane can be
interpreted as the dynamic wormhole. As an example we consider a special
background which leads to the wormhole. In section 4 we speculate about the
wormholes, the number of universes and the observed value of the
cosmological constant $\Lambda \simeq 1.21\times 10^{-52}\left[ 1/m^{2}%
\right] $ in the observed universe. Section 5 is devoted to conclusions.

\section{Wormholes}

In order to obtain a space-time with a wormhole one does a so called surgery
procedure (see [4, 5]). In the four dimensions one starts by taking two
copies $M_{4}$ of the same spacetime and remove a four dimensional region $%
\Omega $ from the one $M_{4}$ and from the second $M_{4}$. As a result one
obtains two spacetimes $M^{+}$ and $M^{-}$. These spacetimes $M^{+}$ and $%
M^{-}$ are joining together along the three-dimensional boundary $\partial
\Omega \equiv \Sigma =\mathbf{R}^{1}\times \Sigma _{2}$, where $\Sigma _{2}$
is a two-dimensional surface, so a new spacetime 
\begin{equation*}
M=M^{+}\cup _{\mathbf{R}^{1}\times \Sigma _{2}}M^{-}
\end{equation*}%
is obtained. In the general case the linear size $\rho $ of $\Sigma _{2}$ is
time-dependent $\rho =\rho \left( \tau \right) $, where $\tau $ is the
proper time on $\Omega $. The spacetime $M$ is the manifold without boundary
that has a \textquotedblleft kink\textquotedblright\ in the geometry at $%
\rho \left( \tau \right) $. The joining can be made in two ways:

1. if one joins together the two external regions $r\in (\rho \left( \tau
\right) ,\infty )$, then the result is a wormhole spacetime $M_{w}$ with two
asymptotic regions.

2. if one joins together the two internal regions $r\in (0,\rho \left( \tau
\right) )$, then the result is a closed baby universe $M_{b}$.

In the case of the wormhole spacetime $M_{w}$ with the spherical symmetry
the surface $\Sigma _{2}$ is two dimensional sphere with a radius $\rho $
and $\Sigma _{2}$ satisfies the flare-out condition: in both sides of $M_{4}$%
, surfaces of constant $r$ increase their areas as one moves away from $%
\Sigma _{2}$; thus one says that $M_{w}$ has a throat at $r=\rho $, which
can be time-dependent. So the topology of the dynamic wormhole $\mathcal{S}$
is:%
\begin{equation*}
\mathcal{S}\approx \mathbf{R}^{1}\times \Sigma _{2}.
\end{equation*}%
As is well-known the spherically symmetric Lorentzian wormhole, a la Morris
and Thorne [4], is defined through the specification of two arbitrary
functions $b\left( r\right) $ and $\varphi \left( r\right) $ with a line
element $ds^{2}=g_{\mu \nu }dx^{\mu }dx^{\nu }$:%
\begin{equation}
ds^{2}=-e^{2\varphi \left( r\right) }dt^{2}+\frac{1}{1-b\left( r\right) /r}%
dr^{2}+r^{2}(d\theta ^{2}+\sin ^{2}\theta d\phi ^{2}).  \tag{2.1}
\end{equation}

Hence the evolution of the wormhole $\mathcal{S}$ is determined by the
Einstein equations projected on $\mathbf{R}^{1}\times \Sigma _{2}$, that is,
by the Lanczos equations:%
\begin{equation}
-\left[ K_{\alpha \beta }\right] +\left[ K\right] \widetilde{g}_{\alpha
\beta }=8\pi GS_{\alpha \beta },  \tag{2.2}
\end{equation}%
where on the left hand of the equation the symbols have meaning: $\widetilde{%
g}_{\alpha \beta }$ is the metric on $\mathcal{S}$, the bracket $[K_{\alpha
\beta }]$ denotes the jump: $[K_{\alpha \beta }]=K_{\alpha \beta
}^{+}-K_{\alpha \beta }^{-}$ across the hypersurface $\mathbf{R}^{1}\times
\Sigma _{2}$ and $\left[ K\right] =\widetilde{g}^{\alpha \beta }\left[
K_{\alpha \beta }\right] $, where $K_{\alpha \beta }^{\pm }$ is the
extrinsic curvature tensor of $\mathcal{S}$ in $M^{\pm }$. On the right side
of the eq. (2.2) $S_{\alpha \beta }$ is the surface stress-energy tensor on $%
\mathcal{S}$.

In the special case the metric (2.1) takes the form:%
\begin{equation}
ds^{2}=-V^{2}\left( r\right) dt^{2}+V^{-2}\left( r\right)
dr^{2}+r^{2}(d\theta ^{2}+\sin ^{2}\theta d\phi ^{2})  \tag{2.3}
\end{equation}%
and the wormhole $\mathcal{S}$ is the 3-dimensional spacetime with the
topology: $\mathbf{R}^{1}\times S^{2}$ which is embedded in $M_{4}$ as
follows:%
\begin{equation}
X:\mathbf{R}^{1}\times S^{2}\rightarrow \mathcal{S}\subset M_{4}  \tag{2.4}
\end{equation}%
where the embedding $X$ is:%
\begin{equation}
X\left( \tau ,\theta ,\phi \right) =\left( t\left( \tau \right) ,\rho \left(
\tau \right) ,\theta ,\phi \right) \in M_{4},  \tag{2.5}
\end{equation}%
the time-like coordinate $t$ and the radius $\rho $ of the sphere $S^{2}$
depend on proper time $\tau $ and $\left( \theta ,\phi \right) \in S^{2}$.
The induced metric $\widetilde{g}_{\alpha \beta }=\partial _{\alpha }X^{\mu
}\partial _{\beta }X^{\nu }g_{\mu \nu }$ on $\mathcal{S}$ is (where $\alpha
=\tau ,\theta ,\phi $):%
\begin{equation}
\widetilde{g}_{\tau \tau }=-\overset{\cdot }{t}^{2}V^{2}\left( \rho \left(
\tau \right) \right) +\overset{\cdot }{\rho }^{2}V^{-2}\left( \rho \left(
\tau \right) \right) ,  \tag{2.6a}
\end{equation}%
\begin{equation}
\widetilde{g}_{\theta \theta }=\rho ^{2}\left( \tau \right) ,\text{ \ }%
\widetilde{g}_{\phi \phi }=\rho ^{2}\left( \tau \right) \sin ^{2}\theta 
\tag{2.6b}
\end{equation}%
and "dot" means differentiation on $\tau $ e.g. $\overset{\cdot }{\rho }%
=d\rho /d\tau $. The manifold $\mathcal{S}$ has three unit tangent vectors $%
T_{\left( a\right) }=\partial _{\alpha }X^{\mu }\partial _{\mu }$: one is
time-like: $T_{\left( \tau \right) }=\partial _{\tau }X^{\mu }\partial _{\mu
}$ :%
\begin{gather}
T_{\left( \tau \right) }=\frac{1}{V^{2}\left( \rho \right) }\sqrt{%
V^{2}\left( \rho \right) +\overset{\cdot }{\rho }^{2}}\frac{\partial }{%
\partial t}+\overset{\cdot }{\rho }\frac{\partial }{\partial r},\text{ } 
\notag \\
T_{\left( \tau \right) }\cdot T_{\left( \tau \right) }=-1,  \tag{2.7}
\end{gather}%
and other ones are space-like:%
\begin{gather}
T_{\left( \theta \right) }=\frac{1}{\rho }\frac{\partial }{\partial \theta },%
\text{ \ }T_{\left( \phi \right) }=\frac{1}{\rho \sin \theta }\frac{\partial 
}{\partial \phi },  \notag \\
T_{\left( \theta \right) }\cdot T_{\left( \theta \right) }=T_{\left( \phi
\right) }\cdot T_{\left( \phi \right) }=+1.  \tag{2.8}
\end{gather}%
The condition $T_{\left( \tau \right) }\cdot T_{\left( \tau \right) }=-1$
leads to that $\overset{\cdot }{t}^{2}=V^{-4}\left( V^{2}+\overset{\cdot }{%
\rho }^{2}\right) $ so the induced metric $\widetilde{g}_{\alpha \beta }$ is:%
\begin{equation}
d\widetilde{s}^{2}=\widetilde{g}_{\alpha \beta }dy^{\alpha }dy^{\beta
}=-d\tau ^{2}+\rho ^{2}\left( \tau \right) \left( d\theta ^{2}+\sin
^{2}\theta d\phi ^{2}\right) .  \tag{2.9}
\end{equation}%
There are two unit normal space-like vectors $n_{\pm }=\pm n^{\mu }\partial
_{\mu }$ to $\mathcal{S}$, one is inward ( belongs to $M^{+}$) and second is
outward ( belongs to $M^{-}$): 
\begin{eqnarray}
n_{\pm } &=&\pm \left( \frac{\overset{\cdot }{\rho }}{V^{2}\left( \rho
\right) }\frac{\partial }{\partial t}+\sqrt{V^{2}\left( \rho \right) +%
\overset{\cdot }{\rho }^{2}}\frac{\partial }{\partial r}\right) ,  \notag \\
n_{\pm }\cdot n_{\pm } &=&+1,\text{ }n_{\pm }\cdot T_{\left( \tau \right)
}=n_{\pm }\cdot T_{\left( \theta \right) }=n_{\pm }\cdot T_{\left( \phi
\right) }=0.  \TCItag{2.10}
\end{eqnarray}%
It means that $\mathcal{S}$\ is the time-like submanifold. In the
orthonormal frame $e_{\widehat{\mu }}^{\mu }$ the metric (2.1) takes the
form:%
\begin{equation}
ds^{2}=e_{\mu }^{\widehat{\mu }}e_{\nu }^{\widehat{\nu }}\eta _{\widehat{\mu 
}\widehat{\nu }}dx^{\mu }dx^{\nu },  \tag{2.11}
\end{equation}%
where the tetrad field $e_{\mu }^{\widehat{\mu }}$ is:%
\begin{equation}
\mathbf{e=}\left( e_{\mu }^{\widehat{\mu }}\right) =diag\left(
V,V^{-1},r,r\sin \theta \right) ,  \tag{2.12}
\end{equation}%
and $\left( \eta _{\widehat{\mu }\widehat{\nu }}\right) =diag\left(
-1,+1,+1,+1\right) $. The inverse tetrad $e_{\widehat{\mu }}^{\mu }$ has the
components: $\mathbf{e}^{-1}\mathbf{=}\left( e_{\widehat{\mu }}^{\mu
}\right) =diag\left( V^{-1},V,r^{-1},r^{-1}\sin ^{-1}\theta \right) $. The
function $V\left( r\right) $ is positive for $r>r_{h}>0$ thus the radius of
the throat $\rho \left( \tau \right) $ of the wormhole is bigger then the
biggest positive root of $V\left( r\right) $. It means that the classical
wormhole has not horizon. The eigenvalues of the operator of the second
fundamental form $b_{n}\left( \cdot \right) =-\nabla _{\cdot }n$ of $%
\mathcal{S}$ give the matrix of the extrinsic curvature $K_{\mu \nu }$ in
the orthonormal frame $K_{\widehat{\mu }\widehat{\nu }}=K_{\mu \nu }e_{%
\widehat{\mu }}^{\mu }e_{\widehat{\nu }}^{\nu }$. Since in our case are two
normal vectors $n_{\pm }$ thus we get two extrinsic curvature $K_{\widehat{a}%
\widehat{a}}^{\pm }$ jumping across $S^{2}$:%
\begin{equation}
b_{n_{\pm }}\left( T_{\left( a\right) }\right) =K_{\widehat{a}\widehat{a}%
}^{\pm }T_{\left( a\right) },  \tag{2.13}
\end{equation}%
where $a=\tau ,\theta ,\phi $. In order to obtain $K_{\widehat{\tau }%
\widehat{\tau }}^{\pm }$ one can notice that for the metric (2.2) there is
the time-like Killing vector $V=\partial _{t}$ and the four-vector
acceleration is proportional to the normal vector $n:\ \nabla _{T_{\left(
\tau \right) }}T_{\left( \tau \right) }=An$. Thus on $\mathcal{S}$ one gets :%
\begin{equation}
K_{\widehat{\tau }\widehat{\tau }}^{\pm }=-A=-\frac{1}{V^{2}n_{\pm }^{t}}%
\frac{d}{d\tau }\left( V^{2}T_{\left( \tau \right) }^{t}\right) .  \tag{2.14}
\end{equation}%
For the above tangent and normal vectors the extrinsic curvature $K_{%
\widehat{a}\widehat{b}}^{\pm }$ is:%
\begin{equation}
K_{\widehat{\tau }\widehat{\tau }}^{\pm }=\mp \frac{1}{\overset{\cdot }{\rho 
}}\frac{d}{d\tau }\left( \sqrt{V^{2}\left( \rho \right) +\overset{\cdot }{%
\rho }^{2}}\right) =\mp \frac{\frac{1}{2}\left. \partial
_{r}V^{2}\right\vert _{r=\rho }+\overset{\cdot \cdot }{\rho }}{\sqrt{%
V^{2}\left( \rho \right) +\overset{\cdot }{\rho }^{2}}},  \tag{2.15}
\end{equation}%
\begin{equation}
K_{\widehat{\theta }\widehat{\theta }}^{\pm }=K_{\widehat{\phi }\widehat{%
\phi }}^{\pm }=\pm \frac{1}{\rho }\sqrt{V^{2}\left( \rho \right) +\overset{%
\cdot }{\rho }^{2}}.  \tag{2.16}
\end{equation}%
One can rewrite $K_{\widehat{\tau }\widehat{\tau }}^{\pm }$ as follows:%
\begin{equation}
K_{\widehat{\tau }\widehat{\tau }}^{\pm }=\mp \left[ \frac{d}{d\tau }\sinh
^{-1}\left( \frac{\overset{\cdot }{\rho }}{V\left( \rho \right) }\right) +%
\frac{1}{2}\partial _{\rho }V^{2}\frac{dt}{d\tau }\right]  \tag{2.17}
\end{equation}%
where $dt/d\tau =V^{-2}\left( \rho \right) \sqrt{V^{2}\left( \rho \right) +%
\overset{\cdot }{\rho }^{2}}$ and $\sinh ^{-1}\left( x\right) =\ln \left( x+%
\sqrt{1+x^{2}}\right) $. The matrix $K$ of the extrinsic curvature $K_{%
\widehat{\mu }\widehat{\nu }}$ in the orthonormal frame is:%
\begin{equation}
K^{\pm }=\left( 
\begin{array}{ccc}
K_{\widehat{\tau }\widehat{\tau }}^{\pm } & 0 & 0 \\ 
0 & K_{\widehat{\theta }\widehat{\theta }}^{\pm } & 0 \\ 
0 & 0 & K_{\widehat{\phi }\widehat{\phi }}^{\pm }%
\end{array}%
\right) .  \tag{2.18}
\end{equation}%
The surface stress-energy tensor $S_{\alpha \beta }$ (for a perfect fluid)
in the orthonormal frame is:%
\begin{equation}
\left( S_{\widehat{a}\widehat{b}}\right) =diag\left( \sigma ,\eta ,\eta
\right) ,  \tag{2.19}
\end{equation}%
where $\sigma $ and $\eta $ are an energy density and a surface tangential
pressure, localized on $\mathcal{S}$, respectively.\ Hence the Lanczos
equations (2.2) take the form:%
\begin{equation}
\frac{1}{\rho }\sqrt{V^{2}\left( \rho \right) +\overset{\cdot }{\rho }^{2}}%
=2\pi G\sigma  \tag{2.20}
\end{equation}%
\begin{equation}
\frac{d}{d\tau }\left( \rho \sqrt{V^{2}\left( \rho \right) +\overset{\cdot }{%
\rho }^{2}}\right) =2\pi G\eta \frac{d\left( \rho ^{2}\right) }{d\tau }. 
\tag{2.21}
\end{equation}%
From these equations one gets relation:%
\begin{equation}
\sigma +\frac{\rho }{2}\frac{d\sigma }{d\rho }=\eta .  \tag{2.22}
\end{equation}%
From the other side the above equations are obtained from the
Hilbert-Einstein action using the thin shell formalism [5-7]. In this
formalism the Riemann tensor in the vicinity of the thin shell given by the
equation $W\left( x\right) =0$ has the form:%
\begin{eqnarray*}
R_{\mu \nu \rho \sigma } &=&-\left( \left[ K_{\mu \rho }\right] n_{\nu
}n_{\sigma }-\left[ K_{\mu \sigma }\right] n_{\nu }n_{\rho }+\left[ K_{\nu
\sigma }\right] n_{\mu }n_{\rho }-\left[ K_{\nu \rho }\right] n_{\mu
}n_{\sigma }\right) \delta \left( W\right) + \\
&&\theta \left( W\right) R_{\mu \nu \rho \sigma }^{+}+\theta \left(
-W\right) R_{\mu \nu \rho \sigma }^{-}.
\end{eqnarray*}%
Hence the Hilbert-Einstein action with a matter field $\Phi $ gives:%
\begin{eqnarray}
S\left[ g,\Lambda ,\Phi \right] &=&\sum_{a=\pm }\int_{M_{\left( a\right)
}}d^{4}x_{\left( a\right) }\sqrt{-g_{\left( a\right) }}\left[ \frac{1}{%
2\kappa }\left( R^{\left( a\right) }\left( g\right) -2\Lambda ^{\left(
a\right) }\right) +\emph{L}_{\left( a\right) }\left( \Phi _{\left( a\right)
},g_{\left( a\right) }\right) \right] +  \notag \\
&&+\frac{c^{2}}{8\pi G}\int_{\partial M=\left\{ W\left( x\right) =0\right\}
}d^{3}x\sqrt{\widetilde{g}^{\left( 3\right) }}Tr\left( \left[ K\right]
\right) .  \TCItag{2.23}
\end{eqnarray}%
In our case $\partial M=\mathcal{S}$ and $R^{\left( a\right) }\left(
g\right) -4\Lambda ^{\left( a\right) }=-\kappa T^{\left( a\right) }$ thus
the above action is:%
\begin{gather}
S\left[ g,\Lambda ,\Phi \right] =\sum_{a=\pm }\int_{M_{\left( a\right)
}}d^{4}x_{\left( a\right) }\sqrt{-g_{\left( a\right) }}\left[ \frac{1}{%
\kappa }\Lambda ^{\left( a\right) }-\frac{1}{2}T^{\left( a\right) }+\emph{L}%
_{\left( a\right) }\left( \Phi _{\left( a\right) },g_{\left( a\right)
}\right) \right] +  \notag \\
+\frac{c^{2}}{4G}\int d\tau \rho ^{2}\left( \tau \right) \left[ 2\frac{d}{%
d\tau }\sinh ^{-1}\left( \frac{\overset{\cdot }{\rho }}{V\left( \rho \right) 
}\right) +\partial _{\rho }V^{2}\frac{dt}{d\tau }+\frac{4}{\rho }\sqrt{%
V^{2}\left( \rho \right) +\overset{\cdot }{\rho }^{2}}\right]  \tag{2.24}
\end{gather}%
where $1/\left( 2\kappa \right) =c^{4}/\left( 16\pi G\right) $ and $T$ is
the trace of the energy-momentum tensor for the matter field: $T=g^{\mu \nu
}T_{\mu \nu }$. The boundary term integrated by parts\ gives:%
\begin{equation}
S\left[ g,\Lambda ,\Phi \right] =S_{m}-\frac{c^{2}}{G}\int d\tau \left[ \rho 
\overset{\cdot }{\rho }\sinh ^{-1}\left( \frac{\overset{\cdot }{\rho }}{%
V\left( \rho \right) }\right) -\rho \sqrt{V^{2}\left( \rho \right) +\overset{%
\cdot }{\rho }^{2}}-\frac{1}{4}\partial _{\rho }V^{2}\frac{dt}{d\tau }\right]
.  \tag{2.25}
\end{equation}%
The last two terms can be rewritten as follows:%
\begin{equation}
\rho \sqrt{V^{2}\left( \rho \right) +\overset{\cdot }{\rho }^{2}}+\frac{1}{4}%
\partial _{\rho }V^{2}\frac{dt}{d\tau }=\left( \rho +\frac{1}{2V^{2}}%
\partial _{\rho }V^{2}\right) \sqrt{V^{2}\left( \rho \right) +\overset{\cdot 
}{\rho }^{2}}.  \tag{2.26}
\end{equation}%
The special form of $V$ is considered in [8].

In the next section we will be interpreting the wormhole as a D2 brane with
the Dirac-Born-Infeld (DBI) action in the background fields of the type IIA
string theory. It means that D2 brane connects two copies of the four
dimensional spacetime $M_{4}$. From the other side the consistent spacetimes
obtained in the type IIA are Minkowski or anti-de Sitter. Thus the form of $%
V^{2}$ is: $V^{2}=+1$ for Minkowski and $V^{2}\left( r\right) =1+\frac{%
\left\vert \Lambda \right\vert }{3}r^{2}$ for anti-de-Sitter.

\section{DBI action for D2-brane and wormhole}

The background fields of Type IIA consist of: the metric $g_{MN},$ the
two-form $B=B_{MN}dX^{M}\wedge dX^{N}$ and the dilaton $\Phi $ and the gauge
fields $C_{\left( i\right) }$ (they are $i$-forms) with$\ i=1,3$. These
gauge fields are coupled to the D2 and D4 branes. The action for the one D2
brane is given by the Dirac-Born-Infeld (DBI) action:%
\begin{eqnarray}
S &=&-T_{2}\int_{\mathcal{M}_{3}}e^{-\Phi }\left( -\det \left( \gamma
_{\alpha \beta }+2\pi \alpha ^{\prime }F_{\alpha \beta }+b_{\alpha \beta
}\right) \right) ^{1/2}d^{3}\xi +  \notag \\
&&T_{2}\int_{\mathcal{M}_{3}}\sum\limits_{i}c_{\left( i\right) }\wedge \exp
\left( 2\pi \alpha ^{\prime }F+b\right) ,  \TCItag{3.1}
\end{eqnarray}%
where $\mathcal{M}_{3}$ is the world-volume of D2-brane embedded into a
background manifold $M_{10}$:%
\begin{equation}
X:\mathcal{M}_{3}\rightarrow M_{10}.  \tag{3.2}
\end{equation}%
The signature of the metric $g_{MN}$ is $\left( -,+,...+\right) $. All the
fields in the DBI action are the pull-back of the background fields by the
embedding $X$:%
\begin{equation}
\gamma =X^{\ast }\left( g\right) ,\text{ }b=X^{\ast }\left( B\right) ,\text{ 
}c_{\left( i\right) }=X^{\ast }\left( C_{\left( i\right) }\right)  \tag{3.3}
\end{equation}%
except for the abelian field $F_{\alpha \beta }$ which is the gauge field on
the D2-brane. Hence the Wess-Zumino (WZ) term is:%
\begin{equation}
\int_{\mathcal{M}_{3}}\sum\limits_{i}c_{\left( i\right) }\wedge \exp \left(
2\pi \alpha ^{\prime }F+b\right) =\int_{\mathcal{M}_{3}}\left[ c_{\left(
1\right) }\wedge \left( 2\pi \alpha ^{\prime }F+b\right) +c_{\left( 3\right)
}\right] .  \tag{3.4}
\end{equation}%
The term $c_{\left( 1\right) }\wedge \left( 2\pi \alpha ^{\prime }F+b\right) 
$ has the form of the Chern-Simons term.\ Assuming that D2-brane is
spherically symmetric with the topology $\mathbf{R}^{1}\times S^{2}$ the
embedding $X$ takes the form:%
\begin{equation}
X\left( \tau ,\theta ,\phi \right) =\left( t\left( \tau \right) ,r\left(
\tau \right) ,\theta ,\phi ,X_{0}^{4},...,X_{0}^{9}\right) ,  \tag{3.5}
\end{equation}%
where $\tau \in \mathbf{R}^{1}$. Thus the fields on the world volume have
the form:%
\begin{equation}
\gamma _{\tau \tau }=\overset{\cdot }{t}^{2}g_{tt}+\overset{\cdot }{r}%
^{2}g_{rr},\text{ \ }\gamma _{\theta \theta }=g_{\theta \theta },\text{ \ }%
\gamma _{\phi \phi }=g_{\phi \phi },  \tag{3.6}
\end{equation}%
\begin{equation}
b_{\tau \theta }=\overset{\cdot }{t}B_{t\theta }+\overset{\cdot }{r}%
B_{r\theta },\text{ \ }b_{\tau \phi }=\overset{\cdot }{t}B_{t\phi }+\overset{%
\cdot }{r}B_{r\phi },\text{ \ }b_{\theta \phi }=B_{\theta \phi },  \tag{3.7}
\end{equation}%
\begin{equation}
c_{\left( 1\right) }=\left( C_{\left( 1\right) t}\overset{\cdot }{t}%
+C_{\left( 1\right) r}\overset{\cdot }{r}\right) d\tau +C_{\left( 1\right)
\theta }d\theta +C_{\left( 1\right) \phi }d\phi ,  \tag{3.8}
\end{equation}%
\begin{equation}
c_{\left( 3\right) }=\left( C_{\left( 3\right) t\theta \phi }\overset{\cdot }%
{t}+C_{\left( 3\right) r\theta \phi }\overset{\cdot }{r}\right) d\tau \wedge
d\theta \wedge d\phi ,  \tag{3.9}
\end{equation}%
\begin{equation}
F=F_{\tau \theta }d\tau \wedge d\theta +F_{\tau \phi }d\tau \wedge d\phi
+F_{\theta \phi }d\theta \wedge d\phi .  \tag{3.10}
\end{equation}%
Here we are looking for a such background for D2 brane which will produce
the action obtained from the Hilbert-Einstein action and leads to the
equation of motion for the throat radius $\rho .$ In the Appendix the
explicit form of DBI action is presented. In order to obtain the action
(2.25) one needs to put constraints on the background fields. Since in this
action no terms linear in $\overset{\cdot }{\rho }$ and $\overset{\cdot }{t}$
nor $\overset{\cdot }{\rho }\overset{\cdot }{t}$ thus one can see that (see
Appendix):%
\begin{equation}
\chi _{t}=\chi _{r}=\Sigma _{tr}=0.  \tag{3.11}
\end{equation}%
These system of equations (3.11) has one simple solution: $B_{t\phi
}=B_{r\phi }=B_{t\theta }=B_{r\theta }=0$. Hence the DBI action takes the
form:%
\begin{gather}
S=-2\pi T_{2}\int_{R^{1}\times \left[ 0,\pi \right] }d\tau d\theta e^{-\Phi }%
\left[ \overset{\cdot }{t}^{2}\left( 1+A^{2}\right) \left\vert
g_{tt}\right\vert -\overset{\cdot }{r}^{2}\left( 1+A^{2}\right) g_{rr}-f^{2}%
\right] ^{1/2}+  \notag \\
+2\pi T_{2}\int_{R^{1}\times \left[ 0,\pi \right] }\left[ \overset{\cdot }{t}%
\Psi _{t}+\overset{\cdot }{r}\Psi _{r}+\Psi _{F}\right] d\tau d\theta , 
\tag{3.12}
\end{gather}%
where $A,f$ and $\Psi _{t},\Psi _{r},\Psi _{F}$ are given in the Appendix.
If one puts the gauge $\gamma _{\tau \tau }=-1$ in (3.6), than the above
action becomes:%
\begin{gather}
S=-2\pi T_{2}\int_{R^{1}\times \left[ 0,\pi \right] }d\tau d\theta e^{-\Phi }%
\left[ 1+A^{2}-f^{2}\right] ^{1/2}+  \notag \\
+2\pi T_{2}\int_{R^{1}\times \left[ 0,\pi \right] }\left[ \pm \Psi _{t}\sqrt{%
\frac{g_{rr}}{\left\vert g_{tt}\right\vert }}\sqrt{\frac{1}{g_{rr}}+\overset{%
\cdot }{r}^{2}}+\overset{\cdot }{r}\Psi _{r}+\Psi _{F}\right] d\tau d\theta .
\tag{3.13}
\end{gather}%
The signs $\pm $ follow from the two solutions of the equation $\gamma
_{\tau \tau }=-1$ with respect to $\overset{\cdot }{t}$. The sign $+$
corresponds to increasing of the coordinate time $t$. Here we chose the sign 
$+$ but we remember that there is the second solution with decreasing time $%
t $. One can see that the first part of the DBI action can be interpreted as
the lagrangian for the matter on the throat of the four dimensional wormhole
where $A$ corresponds to the "magnetic" component of $F$ and $B$ while $f$
is related to the "electric" part of $F$. The WZ term generates dynamics of
D2 brane described by the one degree of freedom $r$ which corresponds to the
radius of the throat. From the above action with the Lagrangian $L$:%
\begin{equation}
L\left( r,\overset{\cdot }{r}\right) =+\Psi _{t}\sqrt{\frac{g_{rr}}{%
\left\vert g_{tt}\right\vert }}\sqrt{\frac{1}{g_{rr}}+\overset{\cdot }{r}^{2}%
}+\overset{\cdot }{r}\Psi _{r}+\Psi _{F}-e^{-\Phi }\left[ 1+A^{2}-f^{2}%
\right] ^{1/2}  \tag{3.14}
\end{equation}%
we get the momentum $p_{r}$ conjugated to $r$:%
\begin{equation*}
p_{r}=\Psi _{t}\sqrt{\frac{g_{rr}}{\left\vert g_{tt}\right\vert }}\frac{%
\overset{\cdot }{r}^{2}}{\sqrt{\frac{1}{g_{rr}}+\overset{\cdot }{r}^{2}}}%
+\Psi _{r}.
\end{equation*}%
Thus the Hamiltonian $H=\overset{\cdot }{r}p_{r}-L$ expressed in $r$ and $%
\overset{\cdot }{r}$ is:%
\begin{equation}
H\left( r,\overset{\cdot }{r}\right) =-\frac{\Psi _{t}}{\sqrt{\left\vert
g_{tt}\right\vert g_{rr}}}\frac{1}{\sqrt{\frac{1}{g_{rr}}+\overset{\cdot }{r}%
^{2}}}+e^{-\Phi }\left[ 1+A^{2}-f^{2}\right] ^{1/2}-\Psi _{F}.  \tag{3.15}
\end{equation}%
From the other side it is known that in the reparametrization invariant
theories Hamiltonian is the constraint: $H=0$. Hence we obtain the equation:%
\begin{equation}
\sqrt{g_{rr}\left\vert g_{tt}\right\vert }\sqrt{\frac{1}{g_{rr}}+\overset{%
\cdot }{r}^{2}}=\frac{\Psi _{t}e^{\Phi }}{\left[ 1+A^{2}-f^{2}\right]
^{1/2}-\Psi _{F}e^{\Phi }}.  \tag{3.16}
\end{equation}%
This equation is the generalization of the (2.20) where the density of the
energy $\sigma $ is replaced by the energy density $\varepsilon $ on the
D2-brane:%
\begin{equation}
\varepsilon =\frac{\Psi _{t}e^{\Phi }}{\left[ 1+A^{2}-f^{2}\right]
^{1/2}-\Psi _{F}e^{\Phi }}.  \tag{3.17}
\end{equation}%
Because the energy density $\varepsilon $ should be positive we get the
following constraints:%
\begin{eqnarray}
e^{-\Phi }\left[ 1+A^{2}-f^{2}\right] ^{1/2} &>&\Psi _{F},  \notag \\
\Psi _{t} &\geq &0.  \TCItag{3.18}
\end{eqnarray}%
This is the general condition on the background fields in which the D2-brane
can be interpreted as the wormhole in the type IIA.

Next we consider the background produced by the spherically symmetric
Dp-brane [13-15] which leads to the wormhole interpretation of the D2-brane.
The background metric $g_{MN}$ is: 
\begin{eqnarray}
ds^{2} &=&g_{MN}dX^{M}dX^{N}=Z_{p}^{-1/2}\left( r\right) \left( -K\left(
r\right) dt^{2}+d\mathbf{x}^{2}\right)  \notag \\
&&+Z_{p}^{1/2}\left( r\right) \left( K^{-1}\left( r\right)
dr^{2}+r^{2}d\Omega _{8-p}^{2}\right) ,  \TCItag{3.19}
\end{eqnarray}%
where $d\mathbf{x}^{2}=dx_{i}dx^{i}$ and $i=1,...,p$. The metric on a unit
round sphere $S^{8-p}$\ is denoted as: $d\Omega _{8-p}^{2}$. The functions $%
Z_{p}$ and $K$ are:%
\begin{equation}
Z_{p}\left( r\right) =1+\alpha _{p}\left( \frac{r_{p}}{r}\right) ^{7-p},%
\text{ \ }K\left( r\right) =1-\left( \frac{r_{H}}{r}\right) ^{7-p}, 
\tag{3.20}
\end{equation}%
where 
\begin{equation}
\alpha _{p}=\left[ 1+\left( r_{H}/r_{p}\right) ^{2\left( 7-p\right) }/2%
\right] ^{1/2}-\left( r_{H}/r_{p}\right) ^{\left( 7-p\right) }/2  \tag{3.21}
\end{equation}%
and%
\begin{equation}
r_{p}^{7-p}=d_{p}\left( 2\pi \right) ^{p-2}g_{s}N\alpha ^{\prime \left(
7-p\right) /2}  \tag{3.22}
\end{equation}%
with the numerical factor $d_{p}=2^{7-2p}\pi ^{\left( 9-3p\right) /2}\Gamma
\left( \left( 7-p\right) /2\right) $. There is the dilaton $\Phi $:%
\begin{equation}
\exp \left( 2\Phi \right) =g_{s}^{2}Z_{p}^{\left( 3-p\right) /2}  \tag{3.23}
\end{equation}%
and the antisymmetric field $B=0$. The number $N$ is the Dp-brane R--R
charge of the p-form $C_{\left( p+1\right) }:$%
\begin{equation}
C_{\left( p+1\right) }=\frac{1}{g_{s}}\left[ \frac{1}{Z_{p}\left( r\right) }%
-1\right] dX^{0}\wedge dX^{1}\wedge ...\wedge dX^{p}.  \tag{3.24}
\end{equation}%
The background metric has the horizon given by $r_{H}$ related to the ADM
mass of the Dp-brane and the singularity is at $r=0$. Since we are in the
frame of the type IIA the dimensions of Dp-branes are: $p=0,2,4,6$. The only
non vanishing component of $C_{\left( p+1\right) }$ is:%
\begin{equation}
C_{\left( p+1\right) t1...p}=\frac{1}{g_{s}}\left[ \frac{1}{Z_{p}\left(
r\right) }-1\right] .  \tag{3.25}
\end{equation}%
For the D2-brane in the above background we get (because $B=0$)$:$%
\begin{equation}
\Psi _{t}=AC_{\left( 1\right) t}\text{ and }\Psi _{F}=\Psi _{r}=0, 
\tag{3.26}
\end{equation}%
where $A=2\pi \alpha ^{\prime }F_{\theta \phi }$ is the magnetic field on
the D2-brane. Hence one can see that the non-trivial background p-form $%
C_{\left( p\right) }$ is for $p=0$. The metric (3.19) for this background is:%
\begin{equation*}
ds^{2}=-Z_{0}^{-1/2}\left( r\right) K\left( r\right)
dt^{2}+Z_{0}^{1/2}\left( r\right) K^{-1}\left( r\right)
dr^{2}+r^{2}Z_{0}^{1/2}\left( r\right) d\Omega _{8}^{2}.
\end{equation*}%
Thus the action (3.13) takes the form:%
\begin{eqnarray}
S &=&-2\pi ^{2}T_{2}\int_{R^{1}}d\tau e^{-\Phi }\left[ 1+A^{2}-f^{2}\right]
^{1/2}  \notag \\
&&+2\pi ^{2}T_{2}\int_{R^{1}}AC_{\left( 1\right) t}\frac{Z_{0}^{1/2}}{%
K\left( r\right) }\sqrt{\frac{K\left( r\right) }{Z_{0}^{1/2}}+\overset{\cdot 
}{r}^{2}}d\tau ,  \TCItag{3.27}
\end{eqnarray}%
where 
\begin{equation}
C_{\left( 1\right) t}=-\frac{\alpha _{0}r_{0}^{7}}{g_{s}\left( r^{7}+\alpha
_{0}r_{0}^{7}\right) }  \tag{3.28}
\end{equation}%
and $r_{0}^{7}=60\pi ^{3}g_{s}\left( \alpha ^{\prime }\right) ^{7/2}N$. The
expression $\alpha _{0}r_{0}^{7}$ is the function of $r_{H}$ and $N$:%
\begin{equation}
\alpha _{0}r_{0}^{7}=\frac{1}{2}\sqrt{4r_{0}^{14}+r_{H}^{14}}-\frac{1}{2}%
r_{H}^{7}  \tag{3.29}
\end{equation}%
The equation (3.16) makes sense in this background only if $AC_{\left(
1\right) t}>0$. Thus one needs that the magnetic field $A=2\pi \alpha
^{\prime }F_{\theta \phi }$ is negative. We obtain from (3.16) the following
equation on $r$:%
\begin{equation}
\overset{\cdot }{r}^{2}=\frac{1}{\sqrt{1+\frac{\alpha _{0}r_{0}^{7}}{r^{7}}}}%
\left( -1+\frac{r_{H}^{7}}{r^{7}}+\frac{A^{2}\alpha _{0}^{2}r_{0}^{14}}{%
sr^{14}}\right) ,  \tag{3.30}
\end{equation}%
where $s=1+A^{2}-f^{2}$. We will introduce a new dimensionaless variable $x$
as follows: $r^{7}=r_{0}^{7}x$. Hence the equation (3.30) becomes:%
\begin{equation}
\overset{\cdot }{x}^{2}=V\left( x\right) ,  \tag{3.31}
\end{equation}%
where:%
\begin{equation}
V\left( x\right) \equiv \frac{49}{r_{0}^{2}}\frac{\left( x_{+}-x\right)
\left( x-x_{-}\right) }{\sqrt{\alpha _{0}+x}}x^{3/14},  \tag{3.32}
\end{equation}%
and: 
\begin{equation}
x_{\pm }=\frac{-1}{2}\left( \alpha _{0}-\frac{1}{\alpha _{0}}\right) \pm 
\frac{1}{2}\sqrt{\left( \alpha _{0}-\frac{1}{\alpha _{0}}\right) ^{2}+4\frac{%
A^{2}}{s}\alpha _{0}^{2}}.  \tag{3.33}
\end{equation}%
In order to get the above formula we used the equation (3.29) which leads to
the relation: $r_{H}^{7}=r_{0}^{7}\left( 1-\alpha _{0}^{2}\right) /\alpha
_{0}.$ Since $r_{H}\geq 0$ we obtain that $\alpha _{0}\in (0,1]$ and $%
x_{-}<0 $. Thus the equation (3.30) makes sense if $V\left( x\right) \geq 0$%
. In this way we obtained that the evolution of the radius of the throat $r$
corresponds to a particle motion in the potential $-V$.\ From the other side
the background metric (3.19) has the horizon given by $r_{H}$. This horizon
corresponds to $x_{H}=\left( 1/\alpha _{0}\right) -\alpha _{0}>0$ and $%
x_{+}>x_{H}$. Thus the wormhole would be visible for a observer at infinity
if the motion is confined to the interval: $(x_{H},x_{+}]$. The equation
(3.31) is highly complicated. In order to get some prediction concerning to
behavior of $x$ we approximate $V$ by a square polynomial $v\left( x\right)
=ax^{2}+bx+c$. The coefficients $a,b$ and $c$ are obtained from the
relations:%
\begin{equation}
v\left( x_{m}\right) =V\left( x_{m}\right) ,\text{ \ }v^{\prime }\left(
x_{m}\right) =0\text{ and }v^{\prime \prime }\left( x_{m}\right) =V^{\prime
\prime }\left( x_{m}\right) ,  \tag{3.34}
\end{equation}%
where $x_{m}$ is the maximum of the function: $V^{\prime }\left(
x_{m}\right) =0$ and $V^{\prime \prime }\left( x_{m}\right) <0$. Hence we
obtain:%
\begin{equation}
v\left( x\right) =\frac{1}{2}V_{m}^{\prime \prime }x^{2}-V_{m}^{\prime
\prime }x_{m}x+\left( V_{m}+\frac{1}{2}V_{m}^{\prime \prime
}x_{m}^{2}\right) ,  \tag{3.35}
\end{equation}%
where $V_{m}^{\prime \prime }=V^{\prime \prime }\left( x_{m}\right) $ and $%
V_{m}=V\left( x_{m}\right) $. Moreover we demand that:%
\begin{equation*}
V\left( 0\right) =v\left( 0\right) =0\text{ and }V\left( x_{+}\right)
=v\left( x_{+}\right) =0.
\end{equation*}%
so: $x_{m}=x_{+}/2$ and $V_{m}^{\prime \prime }=-8V_{m}/x_{+}^{2}<0$ since $%
V_{m}>0$. Thus in the above approximation the equation (3.31) reads:%
\begin{equation}
\overset{\cdot }{x}^{2}=V_{m}\left[ 1-\frac{4}{x_{+}^{2}}\left( x-\frac{x_{+}%
}{2}\right) ^{2}\right] ,  \tag{3.36}
\end{equation}%
where:%
\begin{equation}
V_{m}=\frac{49}{4r_{0}^{2}}\frac{x_{+}\left( x_{+}-2x_{-}\right) }{\sqrt{%
\alpha _{0}+\frac{x_{+}}{2}}}\left( \frac{x_{+}}{2}\right) ^{3/14}>0 
\tag{3.37}
\end{equation}%
The picture of the function $V$ (for $\alpha _{0}=1/2$ and $A^{2}/s=1$)
presented by the solid black line and the quadratic polynomial $v\left(
x\right) =4V_{m}\left( x_{+}-x\right) x/x_{+}^{2}$ (where $x_{+}=\left( 3+%
\sqrt{13}\right) /4$ and $4V_{m}/x_{+}^{2}\simeq 0.984$) presented by the
dots red line is showed below:\FRAME{dtbpFX}{4.4996in}{3in}{0pt}{}{}{Plot}{%
\special{language "Scientific Word";type "MAPLEPLOT";width 4.4996in;height
3in;depth 0pt;display "USEDEF";plot_snapshots TRUE;mustRecompute
FALSE;lastEngine "MuPAD";xmin "0";xmax "2";xviewmin "-0.3023";xviewmax
"2.0023";yviewmin "-0.687439876401282";yviewmax "0.68640626102268";plottype
4;labeloverrides 3;x-label "x";y-label "-V";numpoints 100;plotstyle
"patch";axesstyle "normal";xis \TEXUX{x};yis \TEXUX{y};var1name
\TEXUX{$x$};var2name \TEXUX{$y$};function \TEXUX{$\frac{\left(
3+\sqrt{13}-4x\right) \left( 4x-3+\sqrt{13}\right)
}{16\sqrt{\frac{1}{2}+x}}\root{14}\of{x^{3}}$};linecolor "black";linestyle
1;pointstyle "point";linethickness 1;lineAttributes "Solid";var1range
"0,2";num-x-gridlines 100;curveColor "[flat::RGB:0000000000]";curveStyle
"Line";rangeset"X";function \TEXUX{$0.984\left( 3+\sqrt{13}-4x\right)
x/4$};linecolor "blue";linestyle 3;pointstyle "point";linethickness
1;lineAttributes "Dots";var1range "0,2";num-x-gridlines 100;curveColor
"[flat::RGB:0x000000ff]";curveStyle "Line";rangeset"X";function
\TEXUX{$\frac{-\left( 3+\sqrt{13}-4x\right) \left( 4x-3+\sqrt{13}\right)
}{16\sqrt{\frac{1}{2}+x}}\root{14}\of{-x^{3}}$};linecolor "black";linestyle
1;pointstyle "point";linethickness 1;lineAttributes "Solid";var1range
"-0.3,0";num-x-gridlines 100;curveColor "[flat::RGB:0000000000]";curveStyle
"Line";rangeset"X";valid_file "T";tempfilename
'O3MJEP00.wmf';tempfile-properties "XPR";}} Location of the horizon
corresponds to $x_{H}=3/2$.

Thus in this quadratic approximation the equation (3.36) with the initial
condtion: $x\left( 0\right) =x_{H}$ has the solution:%
\begin{equation}
x\left( \tau \right) =x_{+}\cos ^{2}\left[ \tau \sqrt{\frac{V_{m}}{2x_{+}}}%
-\phi _{0}\right] ,  \tag{3.38}
\end{equation}%
where:%
\begin{equation}
\phi _{0}=\arccos \sqrt{\frac{x_{H}}{x_{+}}}<\pi  \tag{3.39}
\end{equation}%
and $0<x_{H}/x_{+}<1$. In this way we obtained the oscillating whormhole,
which appeares\ on the horizon $x_{H}$ at proper time $\tau =0$ next
reaching maximal size $x_{M}=x_{+}$ at $\tau _{M}:$%
\begin{equation}
\tau _{M}=\sqrt{\frac{2x_{+}}{V_{m}}}\left( \pi -\arccos \sqrt{\frac{x_{H}}{%
x_{+}}}\right)  \tag{3.40}
\end{equation}%
and vanishing under the horizon $x_{H}$ at time $\tau _{f}$:%
\begin{equation}
\tau _{f}=2\sqrt{\frac{2x_{+}}{V_{m}}}\arccos \sqrt{\frac{x_{H}}{x_{+}}}. 
\tag{3.41}
\end{equation}%
Hence the evolution of the throat radius $r$ is confined to the domain:%
\begin{equation}
r_{+}\geq r>r_{H},  \tag{3.42}
\end{equation}%
where: $r_{+}=r_{0}\left( x_{+}\right) ^{-1/7}$ and:%
\begin{equation}
r\left( \tau \right) =r_{0}\sqrt[7]{x\left( \tau \right) }.  \tag{3.43}
\end{equation}%
The requirement $r>r_{H}$ is implicated by the fact that $r_{H}$ is the
horizon in the background metric (3.19). In otherwise the D2 wormhole has
the throat under the horizon (with the radius $r_{H}$) of the background
metric and becomes "invisible" for an observer at infinity. This condition
is accomplished by the above solution. Hence we obtained in the IIA string
theory the transient wormhole with the time dependent throat $r$ given by
(3.43).

\section{Vacuum energy and wormholes}

The vacuum energy obtained form the quantum field theory (QFT) depends on
the energy scale up to which we trust the theory. The cutoffs of the energy
changing from the order of supersymmetry breaking scale (which is probably
of the order 10-100TeV or bigger) to the Planck scale. Hence the vacuum
energy density is changing from $10^{-64}$\ to $1$\ in the Planck units.

From the other side the observation of the apparent luminosity of distant
supernovae [16, 17] indicates that the expansion of the universe has
recently begun to accelerate. In order to explains this unexpected
acceleration one introduced so called "dark energy". The observation leads
to value of the dark energy density which is equal to $\left( 1.35\pm
0.15\right) \times 10^{-123}$.\ Next this energy density is identify with
the cosmological constant $\Lambda $. Expressing $\Lambda $\ in meter$^{-2}$%
\ one obtains the value $\Lambda \simeq 1.21\times 10^{-52}\left[ 1/m^{2}%
\right] $.\ Moreover as turn outs the dark energy is experimentally
indistinguishable from vacuum energy. But as one can see the theoretical
vacuum energy (obtained from the quantum field theory)\ is in huge
discrepancy with the observed dark energy represented by the cosmological
constant. \ 

We propose a mechanism for an explanation of this discrepancy. This
mechanism is based on the assumption that there are wormholes of the Planck
size and that the vacuum energy is flowed across these wormholes. But not
all energy can be drained by the wormholes. Since the throats $r_{w}$ of
wormholes are the size of the Planck length $l_{Pl}$ the energy which can be
flowed by the one wormhole is: $E_{w}=\hbar c/r_{w}$. Hence we get the
following equation:%
\begin{equation}
E_{c}-E_{o}=NE_{w},  \tag{4.1}
\end{equation}%
where $E_{c}$ is the vacuum energy in the volume $V$ obtained from QFT, $%
E_{o}$ is the vacuum energy in this same volume $V$ obtained from the
cosmological observations: $E_{o}=\Lambda Vc^{4}/\left( 4\pi G\right) $ and $%
N$ is the number of the wormholes in $V$. Thus the density number $n=N/V$ of
wormholes expressed in units $[1/m^{3}]$ is:%
\begin{equation}
\frac{1}{4\pi }\frac{r_{w}\Lambda }{l_{Pl}^{2}}\left( \frac{E_{c}}{E_{o}}%
-1\right) =n.  \tag{4.2}
\end{equation}%
This number crucially depends on the cutoff of the energy scale via $E_{c}$.
Because the vacum energy density represented by $\Lambda $ is constant the
excces of energy should flowed to the other universes which are connected to
our universe by these wormholes. If we assume that supersymmetry exists,
then the number of other universes will be depend on the fundamental
supersymmetry breaking scale $E_{SUSY}$. It means that the universes with
different values of $E_{SUSY}$ will have different number connected
universes by the wormholes. For example if the breaking scale is of order
100TeV, then: $E_{c}/E_{o}\simeq 10^{60}$. Thus we get that the number
density of the wormholes with the size of throats $l_{Pl}$ is equal to:%
\begin{equation}
n\simeq 10^{42}.  \tag{4.3}
\end{equation}%
The observable universe has volume $10^{80}$ (in cubic meters $m^{3}$) so
the total number $N$ of the wormholes in the observable universe is $%
10^{122} $. Hence the number of other universes connected to our by the
wormholes is: $10^{122}.$ It will be interesting relate this number to the
number of vacua obtained from the string theory landscape.

\section{Conclusions}

The purpose of this paper was relate the four-dimensional lorentzian
wormholes to the backgrounds of string theory. From the dimensional
arguments we get that the appropriate backgrounds are given by the type IIA
string theory. We related DBI action for D2 brane to the action of dynamic
wormhole. This relation between these actions led us to the conditions on
the backgrounds with the D2 branes which can be interpreted as the wormholes
in the four dimensional spacetimes. These spacetimes had to be
asymptotically Minkowski or anti-de Sitter. The background fields acquired
interpretation of the matter on the throat of the wormhole. The dynamics of
the throat is obtained from WZ term in the DBI action. In this way the
four-dimensional wormhole took on the form of D2 brane in the type IIA
supergravity approximation of the string theory. The energy condition on the
matter supporting the wormhole is given by eq. (3.18). As the example we
considered special solution of IIA which produced the transient wormhole. We
also speculated about the relation of number of universes, the value of the
cosmological constant and supesymmetry breaking scale.

\section{Appendix}

\subsection{DBI action}

The determinant of the sum of a symmetric $g=\left( g_{ab}\right) $ and an
antisymmetric $\mathcal{B}=\left( \mathcal{B}_{ab}\right) $ 3x3 matrices is:%
\begin{equation*}
\det \left( g+\mathcal{B}\right) =\det g\det \left( 1+g^{-1}\mathcal{B}%
\right) .
\end{equation*}%
For the diagonal matrix $g=diag\left( g_{1},g_{2},g_{3}\right) $ one gets:%
\begin{equation*}
\det \left( g+\mathcal{B}\right)
=g_{1}g_{2}g_{3}+b_{1}^{2}g_{1}+b_{2}^{2}g_{2}+b_{3}^{2}g_{3},
\end{equation*}%
where $b_{a}=\varepsilon _{abc}\mathcal{B}_{bc}/2$: $b_{1}=\mathcal{B}_{23},$
$b_{2}=-\mathcal{B}_{13},$ $b_{3}=\mathcal{B}_{12}$.

The DBI action for a D2 brane with the background fields (3.6 - 3.10) is:%
\begin{eqnarray*}
S &=&-T_{2}\int_{R^{1}\times S^{2}}d\tau d\theta d\phi e^{-\Phi }\left[ 
\overset{\cdot }{t}^{2}\Sigma _{tt}-2\overset{\cdot }{t}\overset{\cdot }{r}%
\Sigma _{tr}-\overset{\cdot }{r}^{2}\Sigma _{rr}-2\overset{\cdot }{t}\chi
_{t}-2\overset{\cdot }{r}\chi _{r}-f^{2}\right] ^{1/2}+ \\
&&T_{2}\int_{R^{1}\times S^{2}}\left[ \overset{\cdot }{t}\Psi _{t}+\overset{%
\cdot }{r}\Psi _{r}+\Psi _{F}\right] d\tau d\theta d\phi ,
\end{eqnarray*}%
where:%
\begin{equation*}
\Sigma _{tt}=\left( 1+A^{2}\right) \left\vert g_{tt}\right\vert -g_{\theta
\theta }B_{t\phi }^{2}-g_{\phi \phi }B_{t\theta }^{2},
\end{equation*}%
\begin{equation*}
\Sigma _{rr}=\left( 1+A^{2}\right) g_{rr}+g_{\theta \theta }B_{r\phi
}^{2}+g_{\phi \phi }B_{r\theta }^{2},
\end{equation*}%
\begin{equation*}
\Sigma _{tr}=g_{\theta \theta }B_{t\phi }B_{r\phi }+g_{\phi \phi }B_{t\theta
}B_{r\theta },
\end{equation*}%
\begin{equation*}
\chi _{t}=2\pi \alpha ^{\prime }\left( g_{\theta \theta }B_{t\phi }F_{\tau
\phi }+g_{\phi \phi }B_{t\theta }F_{\tau \theta }\right) ,
\end{equation*}%
\begin{equation*}
\chi _{r}=2\pi \alpha ^{\prime }\left( g_{\theta \theta }B_{r\phi }F_{\tau
\phi }+g_{\phi \phi }B_{r\theta }F_{\tau \theta }\right) ,
\end{equation*}%
\begin{equation*}
f^{2}=\left( 2\pi \alpha ^{\prime }\right) ^{2}\left( g_{\theta \theta
}F_{\tau \phi }^{2}+g_{\phi \phi }F_{\tau \theta }^{2}\right) ,
\end{equation*}%
\begin{equation*}
A=B_{\theta \phi }+2\pi \alpha ^{\prime }F_{\theta \phi }.
\end{equation*}%
The terms in WZ part reads:%
\begin{equation*}
\Psi _{t}=AC_{\left( 1\right) t}-C_{\left( 1\right) \theta }B_{t\phi
}+C_{\left( 1\right) \phi }B_{t\theta }+C_{\left( 3\right) t\theta \phi },
\end{equation*}%
\begin{equation*}
\Psi _{r}=AC_{\left( 1\right) r}-C_{\left( 1\right) \theta }B_{r\phi
}+C_{\left( 1\right) \phi }B_{r\theta }+C_{\left( 3\right) r\theta \phi },
\end{equation*}%
\begin{equation*}
\Psi _{F}=2\pi \alpha ^{\prime }\left( -C_{\left( 1\right) \theta }F_{\tau
\phi }+C_{\left( 1\right) \phi }F_{\tau \theta }\right) .
\end{equation*}

\section{References}

[1] A. Einstein and N. Rosen: \textit{The particle problem in the general
theory of relativity}, Phys. Rev. 48 (1935), 73.

[2] C. Bambi, \textit{Can the supermassive objects at the centers of
galaxies be traversable wormholes? The first test of strong gravity for
mm/sub-mm VLBI facilities, }Phys. Rev.D87 : 107501, 2013, [arXiv:1304.5691];

[3] N. S. Kardashev, I. D. Novikov and A. A. Shatskiy, \textit{Astrophysics
of Wormholes}, Int. J. Mod. Phys. D 16, 909 (2007) [arXiv: astro-ph/0610441].

[4] M.S. Morris and K.S. Thorne, \textit{Wormholes in spacetime and their
use for interstellar travel: A tool for teaching general relativity}, Am. J.
Phys. 56 (1988) 395. \ \ 

[5] M. Visser, \textit{Lorentzian Wormholes: From Einstein to Hawking},
American Institute of Physics, Woodbury, 1995.

[6] W. Israel, \textit{Singular hypersurfaces and thin shells in general
relativity}, Nuovo Cimento B44 (1966) 1.

[7] S. Mukohyama, \textit{Doubly covariant action principle of singular
hypersurfaces in general relativity and scalar-tensor theories}, Phys.Rev.
D65 (2002) 024028 [arXiv: gr-qc/0108048].

[8] C. Barcelo and M. Visser, \textit{Brane surgery: energy conditions,
traversable wormholes, and voids,}\ Nuclear Physics B 584 (2000) 415--435.

[9] G. W. Gibbons, M. B. Green and M. J. Perry, \textit{Instantons and
Seven-Branes in Type IIB Superstring Theory}, Phys. Lett. B370 (1996) 37--44
[arXiv: hep-th/9511080].

[10] E. Bergshoeff, A. Collinucci, U. Gran, D. Roest and S. Vandoren, 
\textit{Non-extremal D-instantons}, JHEP 0410:031, 2004, [arXiv:
hep-th/0406038].

[11] E. Bergshoeff, A. Collinucci, U. Gran, D. Roest and S. Vandoren, 
\textit{Non-extremal instantons and wormholes in string theory},
Fortsch.Phys. 53 (2005) 990-996 [arXiv: hep-th/0412183]

[12] J. Maldacena and L. Maoz, \textit{Wormholes in AdS}, JHEP 0402, 053
(2004), [arXiv: hep-th/0401024].

[13] G. T. Horowitz and A. Strominger, \textit{Black strings and p-branes},
Nucl. Phys. B360, 197 (1991).

[14 M. J. Duff, Ramzi R. Khuri and J. X. Lu, \textit{String Solitons}, Phys.
Rept. 259, 213 (1995), [arXiv: hep-th/9412184].

[15] A. Strominger, \textit{Massless Black Holes and Conifolds in String
Theory}, Nucl. Phys. B451, 96 (1995), [arXiv: hep-th/9504090].

[16] Supernova Search Team Collaboration, A. G. Riess et al., \textit{%
Observational Evidence from Supernovae for an Accelerating Universe and a
Cosmological Constant}, Astron. J. 116 (1998) 1009--1038, [arXiv:
astro-ph/9805201].

[17] Supernova Cosmology Project Collaboration, S. Perlmutter et al., 
\textit{Measurements of Omega and Lambda from 42 High-Redshift Supernovae},
Astrophys. J. 517 (1999) 565--586, [arXiv: astro-ph/9812133].

\end{document}